\documentclass[global,twocolumn]{svjour}
\usepackage[dvips]{graphicx}
\usepackage[dvips]{color} 
\usepackage{bbm}
\usepackage{subfigure}
\usepackage{amsmath}
\usepackage{amssymb}

\newcommand{\trace}[1]{\text{Tr}\left\{ #1\right\}}

\newcommand{\imag}{{\mathrm i}}

\newcommand{\pdiff}[2]{\frac{\partial #1}{\partial #2}}

\renewcommand{\d}{\mathrm d}

\renewcommand{\Im}[1]{\text{Im}\left\{ #1\right\}}
\renewcommand{\Re}[1]{\text{Re}\left\{ #1\right\}}

\renewcommand{\hbar}{\hslash}
\newcommand{\unitmatrix}{\mathbbm{1}}
\journalname{Applied Physics A (2007)}
\begin{document}
\title{Circuit theory for crossed Andreev reflection and nonlocal
  conductance}
\author{Jan Petter Morten\inst{1,3} \and Arne
  Brataas\inst{1,3} \and Wolfgang Belzig\inst{2,3}
}                     % Do not remove
\institute{Department of Physics, Norwegian University of Science and
  Technology, N-7491 Trondheim, Norway \and University of Konstanz,
  Department of Physics, D-78457 Konstanz, Germany \and Centre for
  Advanced Study, Drammensveien 78, N-0271 Oslo, Norway}
\date{Received: 31 January 2007 / Revised version: 12 April 2007}
\maketitle
\begin{abstract}
  Nonlocal currents, in devices where two \\normal-metal terminals are
  contacted to a superconductor, are determined using the circuit
  theory of mesoscopic superconductivity. We calculate the conductance
  associated with crossed Andreev reflection and electron transfer
  between the two normal-metal terminals, in addition to the
  conductance from direct Andreev reflection and quasiparticle
  tunneling. Dephasing and proximity effect are taken into account.
\end{abstract}
\section{\label{sec:intro}Introduction}
\begin{figure}[htbp]
  \centering
  %* Start of pstex figure
  \begin{picture}(0,0)%
\includegraphics{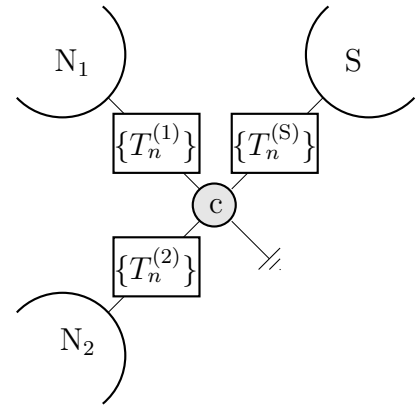}%
\end{picture}%
\setlength{\unitlength}{1776sp}%
\begingroup\makeatletter\ifx\SetFigFont\undefined%
\gdef\SetFigFont#1#2#3#4#5{%
  \reset@font\fontsize{#1}{#2pt}%
  \fontfamily{#3}\fontseries{#4}\fontshape{#5}%
  \selectfont}%
\fi\endgroup%
\begin{picture}(5596,5596)(503,-6759)
\put(1876,-3263){\makebox(0,0)[lb]{\smash{\SetFigFont{5}{6.0}{\familydefault}{\mddefault}{\updefault}{\color[rgb]{0,0,0}{\large $\{T^{(1)}_n\}$}}%
}}}
\put(1876,-4996){\makebox(0,0)[lb]{\smash{\SetFigFont{5}{6.0}{\familydefault}{\mddefault}{\updefault}{\color[rgb]{0,0,0}{\large $\{T^{(2)}_n\}$}}%
}}}
\put(3526,-3271){\makebox(0,0)[lb]{\smash{\SetFigFont{5}{6.0}{\familydefault}{\mddefault}{\updefault}{\color[rgb]{0,0,0}{\large $\{T^{(\text{S})}_n\}$}}%
}}}
\put(1051,-2086){\makebox(0,0)[lb]{\smash{\SetFigFont{5}{6.0}{\familydefault}{\mddefault}{\updefault}{\color[rgb]{0,0,0}{\large N$_1$}}%
}}}
\put(5101,-2086){\makebox(0,0)[lb]{\smash{\SetFigFont{5}{6.0}{\familydefault}{\mddefault}{\updefault}{\color[rgb]{0,0,0}{\large S}}%
}}}
\put(1126,-5986){\makebox(0,0)[lb]{\smash{\SetFigFont{5}{6.0}{\familydefault}{\mddefault}{\updefault}{\color[rgb]{0,0,0}{\large N$_2$}}%
}}}
\put(3211,-4073){\makebox(0,0)[lb]{\smash{\SetFigFont{5}{6.0}{\familydefault}{\mddefault}{\updefault}{\color[rgb]{0,0,0}{\large c}}%
}}}
\end{picture}
  %* End of pstex figure
\caption{Our circuit theory model: A cavity (c) is connected to one
  superconducting (S) and two normal-metal terminals (N$_1$ and
  N$_2$). The three connectors are described by their sets of
  transmission probabilities. A coupling to ground represents the
  ``leakage current'' (see text).}
  \label{fig:ct}
\end{figure}

%*******************************************************************
Transport between a normal-metal and a superconductor at subgap energy
is possible through Andreev reflections, where an incident electron
from the normal-metal is retro-reflected as a hole and a Cooper pair
is transferred into the superconductor \cite{andreef:JETP1228}.
However, since Andreev reflection is a nonlocal process on the scale
of the coherence length, the retro-reflected hole can end up in
another normal-metal contacted to the superconductor. This process is
known as crossed Andreev reflection (CA)
\cite{byers:306,deutsher:apl00} and contributes to a nonlocal
conductance. We define the nonlocal conductance in a three-terminal
device (see Fig. \ref{fig:ct}) as the current response in one
normal-metal terminal (N$_1$) to a voltage bias between another
normal-metal terminal (N$_2$) and a superconducting terminal (S).  The
nonlocal conductance will also depend on electron transfer (ET)
between the two normal-metal terminals, which gives a contribution
with opposite sign as CA to the nonlocal conductance.\footnote{In our
  previous paper Ref.  \cite{morten:prb06} this transport process was
  referred to as electron cotunneling, but as this phrase applies to
  the tunneling limit we will here use the more general term electron
  transfer.} The nonlocal conductance of such systems has recently
been studied extensively both experimentally
\cite{beckmann:prl04,russo:prl05,cadden:237003} and theoretically
\cite{Falci:epl01,yamashita:prb03-174504,sanchez:214501,chtchelkatchev:jetp03,melin:174509,morten:prb06,kalenkov:172503,brinkman:214512,levyyeyati:cond-mat/0612027}
since it demonstrates inherently mesoscopic physics, and since crossed
Andreev reflection is a way to produce spatially separated entangled
electron pairs.

In our recent paper Ref. \cite{morten:prb06}, we used the circuit
theory of mesoscopic superconductivity \cite{nazarov:sm99} to
calculate the conductances in a three terminal device where two
normal-metal terminals and one superconducting terminal are connected
to a region where chaotic scattering takes place. We assumed that that
the energy of the injected particles was much smaller than the gap of
the superconducting terminal. In this case, the only transport process
involving only one of the normal-metal terminals and the
superconductor is direct Andreev reflection (DA), where the hole is
backreflected into the same normal-metal as the incident electron. We
now extend this approach to take into account situations where the
bias voltage is comparable to the gap of the superconductor so that
incident electrons from a normal-metal may be transferred into the
superconductor as quasiparticles (QP). Taking these processes into
account, the current at energy $E$ out of N$_1$ can be written
\begin{align}
  \label{eq:iE}
  I_1(E)=&\;\frac{G_\text{CA}(E)}{e}\left[1-f_1(E)-f_2(-E)\right]\nonumber\\
  &+\frac{G_\text{ET}(E)}{e}\left[f_2(E)-f_1(E)\right]\nonumber\\
  &+2\frac{G_\text{DA}(E)}{e}\left[1-f_1(E)-f_1(-E)\right]\nonumber\\
  &+\frac{G_\text{QP}(E)}{e}\left[f_\text{S}(E)-f_1(E)\right]\,.
\end{align}
Here, the functions $f_n(\pm E)$ denote the Fermi-Dirac distribution
functions in terminals $n=1,2,\text{S}$ at energy $\pm E$ and we have
defined energy dependent conductances $G(E)$ for the transport
processes discussed above. Total charge current is given by
$I_\text{charge,1}=\int\d E I_1(E)$. The nonlocal differential
conductance is obtained from Eq. \eqref{eq:iE},
\begin{align}
  \pdiff{I_\text{charge,1}}{V_2}=-\int\d E \left[
  G_\text{ET}(E)-G_\text{CA}(E) \right]\pdiff{f(E-eV_2)}{E}.
\label{eq:chargecurrent}
\end{align}
This shows how the nonlocal conductance is determined by competing
contributions from crossed Andreev reflections and electron transfer.

\section{\label{sec:circuittheory}Circuit theory}
%*******************************************************************
In the circuit theory of mesoscopic superconductivity
\cite{nazarov:sm99}, a system can be modeled as a network of
terminals, connectors and nodes in a similar manner as electric
circuits on the macroscale are modeled using classical circuit theory.
The theory is formulated in terms of matrix Green's functions and
matrix currents in Nambu-Keldysh space\footnote{In our notation
  $\hat{\sigma}_i$ is a Pauli matrix in Nambu space and $\bar{\tau}_i$
  is a Pauli matrix in Keldysh space.  Matrices with structure in both
  spaces are denoted with accent $\check{ }$.} and {\it e.g.}
describes quantum effects and flow of charge, energy and particle-hole
correlations.

To describe the three-terminal devices at hand we introduce
equilibrium Green's functions depending on temperature and local
chemical potential for the two normal-metal terminals and the
superconducting terminal. The terminals are coupled to a region where
scattering takes place, which we will refer to as a cavity. Our
circuit theory model is shown in Fig. \ref{fig:ct}. If we assume that
the distance between the contacts is very small on the scale of the
coherence length, the properties of the cavity are spatially
homogeneous so that it can be described by one nonequilibrium Green's
function $\check{G}_\text{c}$.  If spatial variation in the scattering
region is important, it can be modeled as a network of cavities with
different Green's functions. We assume that $\check{G}_\text{c}$ is
isotropic due to chaotic or diffusive scattering.

Connectors between terminals and the cavity are described by their
sets of transmission probabilities $\{T^{(n)}\}$, with flow of matrix
currents $\check{I}_n$ depending on the Green's functions of adjacent
elements,
\begin{equation}
  \check{I}_n=-2\frac{e^2}{\pi\hbar}\sum_k T^{(n)}_k\frac{\left[\check{G}_n,\check{G}_\text{c}\right]}{4+T_k^{(n)}\left(\left\{\check{G}_n,\check{G}_\text{c}\right\}-2\right)}.
  \label{eq:matrixcurrent}
\end{equation}
The matrix current describes not only the flow of charge, spin, and
energy currents, but {\it e.g.} also the flow of correlations
\cite{nazarov:sm99}.

Finally, the theory is completed by a generalized ``Kirchhoff's
rule'': The sum of matrix currents flowing into a cavity should
vanish. This determines the nonequilibrium Green's function of the
cavity in our present circuit. The spectral charge current through
connector $n$ can be determined
$I_{\text{T},n}=\trace{\hat{\sigma}_3\hat{I}_n^\text{(K)}}/8e$ once
the Green's functions  $\check{G}_n,\,\check{G}_\text{c}$ have been
determined. Similarly, the spectral energy current becomes
$I_{\text{L},n}=\trace{\hat{I}_n^\text{(K)}}/8e$. The K superscript
denotes the Keldysh matrix block of the current.

Using the generic circuit theory model above, we can describe crossed
Andreev reflection in a wide range of different experimental systems.
For example, consider a system where two adjacent normal-metal
electrodes are deposited on and connected by metallic contacts to a
mesoscopic superconductor. The terminal Green's functions correspond
to properties of the normal-metals and the superconducting sample away
from the contact region. The cavity corresponds to the small region of
the superconductor between the normal-metal contacts.  Metallic
contacts are modeled by putting $T^{(n)}_k=1$ for the conducting modes
of contacts $n=1,2$. The contact between the cavity and the
superconducting terminal can be modeled as a diffusive connector
introducing a bimodal distribution of transmission probabilities
\cite{Belzig:prb00}. Other systems for experimental study of crossed
Andreev reflection may be fabricated from superconductors coupled to
semiconductors where the geometry is defined by deposition of gates.
For example, the semiconductor can consist of a ballistic cavity with
two point contacts to normal reservoirs.  In this case, the cavity
Green's function describes the nonequilibrium state of the ballistic
cavity.  The point contacts are modeled by putting $T^{(n)}_k=1$ for
the open channels and zero otherwise for $n=1,2$.  The transparency
and number of conducting modes of the contact to the superconductor
can also be determined experimentally.

Superconducting pairing and dephasing in the cavity are described in
circuit theory by introducing a ``leakage current''
$\check{I}_\text{leakage}=-\imag e^2\nu_0
V_\text{c}[\check{G}_\text{c},\check{H}_\text{c}]$ in the matrix
current conservation on the cavity \cite{nazarov:sm99}. Here $\nu_0$
is the density of states, $V_\text{c}$ the volume,
$\check{H}_\text{c}=E\hat{\sigma}_3+\imag\hat{\sigma}_1\Re{\Delta_\text{c}}+\imag\hat{\sigma}_2\Im{\Delta_\text{c}}$
the Hamiltonian, and $\Delta_\text{c}$ the gap in the cavity. The
energy-dependent term in $\check{H}_\text{c}$ describes dephasing
between electrons and holes, and sets an energy scale for the
proximity effect which we will refer to as the effective Thouless
energy of the cavity. When we disregard Josephson effects, the phase
of $\Delta_\text{c}$ can be chosen arbitrarily e.g. purely imaginary
and inspection of the retarded part of the matrix current conservation
reveals that pairing inside the cavity appears with the same matrix
structure as the coupling to the superconducting terminal. Therefore,
the effect of pairing inside the cavity can be described by a
renormalization of the coupling between the cavity and the
superconducting terminal. Thus the difference between a normal and a
superconducting cavity is equivalent to rescaling this conductance,
i.e. only quantitative modifications.

\section{\label{sec:results}Results}
%*******************************************************************
We will now discuss the structure of the Green's functions and the
solution of the matrix equations. The normal terminals have Green's
functions
$\check{G}_{1(2)}=\hat{\sigma}_3\bar{\tau}_3+(\hat{\sigma}_3h_{\text{L},1(2)}+\hat{\unitmatrix}h_{\text{T},1(2)})(\bar{\tau}_1+\imag\bar{\tau}_2)$,
where we have introduced the charge- and energy-distribution functions
$h_\text{T}(E),\,h_\text{L}(E)$ that can be written in terms of the
particle distribution function $f(E)$ as $h_\text{T}=1-f(E)-f(-E)$ and
$h_\text{L}=-f(E)+f(-E)$, see Ref. \cite{Belzig:sm99}. The retarded
(advanced) part of the Green's function of the superconducting
reservoir is
$\hat{G}^\text{R(A)}_\text{S}=([E\pm\imag\delta]\hat{\sigma}_3+\Delta\imag\hat{\sigma}_2)/\Omega$,
where $\Omega=\sqrt{(E\pm\imag\delta)^2-\Delta^2}$, $\Delta$ is the
gap. The Keldysh part is obtained from
$\hat{G}^\text{K}_\text{S}=\hat{G}^\text{R}_\text{S}\hat{h}_\text{S}-\hat{h}_\text{S}\hat{G}^\text{A}_\text{S}$
where
$\hat{h}_\text{S}=\hat{\sigma}_3h_\text{L,S}+\hat{\unitmatrix}h_\text{T,S}$.
We parametrize the Green's function of the cavity as
$\hat{G}^\text{R}_\text{c}=\hat{\sigma}_3\cosh(\theta)+\imag\hat{\sigma}_2\sinh(\theta)$
and
$\hat{G}^\text{A}_\text{c}=-\hat{\sigma}_3(\hat{G}^\text{R}_\text{c})^\dag\hat{\sigma}_3$,
the Keldysh part is given by
$\hat{G}^\text{K}_\text{c}=\hat{G}^\text{R}_\text{c}~\hat{h}_\text{c}-\hat{h}_\text{c}~\hat{G}^\text{A}_\text{c},$
where
$\hat{h}_\text{c}=\hat{\unitmatrix}h_\text{L,c}+\hat{\sigma}_3h_\text{T,c}$.

With the Green's functions specified as above, we impose matrix
current conservation in the cavity, $\sum_n
\check{I}_n+\check{I}_\text{leakage}=0$ to obtain equations that
determine $\theta$, $h_\text{T,c}$, and $h_\text{L,c}$. The equations
for the distribution functions are conservation of charge (energy) at
each energy,
\begin{align}
  \sum_n G_{\text{T(L)},n}\left(h_{\text{T(L)},n}-h_\text{T(L),c}\right)=0,
\end{align}
for $n=1,2,\text{S}$, where we have defined effective, energy dependent
conductances for charge (energy) $G_{\text{T(L)},n}[\theta(E)]$
between reservoir $n$ and the cavity. Using the solution for
$\check{G}_\text{c}$ that we have obtained, we calculate the current
out of terminal N$_1$ and compare the result to Eq. \eqref{eq:iE}.
This allows us to determine the conductances for the transport
processes:
\begin{subequations}
  \begin{align}
    G_\text{QP}(E)=&\;\frac{G_\text{L,1}G_\text{L,S}}{G_\text{L,1}+G_\text{L,2}+G_\text{L,S}},\\
    G_\text{DA}(E)=&\;\frac{1}{4}\left(\frac{G_\text{T,1}\left(G_\text{T,2}+G_\text{T,S}\right)}{G_\text{T,1}+G_\text{T,2}+G_\text{T,S}}\right.\nonumber\\
      &\;\left.-\frac{G_\text{L,1}\left(G_\text{L,2}+G_\text{L,S}\right)}{G_\text{L,1}+G_\text{L,2}+G_\text{L,S}}\right),\\
    G_{\stackrel{\scriptstyle{\text{ET}}}{\scriptstyle{\text{CA}}}}(E)=&\;\frac{1}{2}\left(\frac{G_\text{L,1}G_\text{L,2}}{G_\text{L,1}+G_\text{L,2}+G_\text{L,S}}\right.\nonumber\\
      &\;\left.\pm\frac{G_\text{T,1}G_\text{T,2}}{G_\text{T,1}+G_\text{T,2}+G_\text{T,S}}\right).\label{eq:etca}
  \end{align}
  \label{eq:conductances}
\end{subequations}
The conductance for quasiparticle transport into the superconducting
terminal, $G_\text{QP}$, is proportional to $G_\text{L,S}$ which is
the energy conductance for transport from the cavity into the
superconductor. This quantity vanishes at subgap energies where
Andreev reflection of particles from opposite sides of the Fermi
surface is the only possible transport process. The symmetry between
$G_\text{ET}$ and $G_\text{CA}$ in Eq. \eqref{eq:etca} was discussed
in our previous paper Ref.  \cite{morten:prb06}, and shows that the
differential nonlocal conductance, given by $G_\text{ET}-G_\text{CA}$
see Eq.  \eqref{eq:chargecurrent}, is always positive. In that paper,
we also discussed the limit that there is no resistance between the
superconducting reservoir and the cavity. We see from Eq.
\eqref{eq:etca} that in this case $G_\text{ET(CA)}$ vanishes because
of the large terms $G_\text{T,S},\, G_\text{L,S}$.

\begin{figure}[h]
  \includegraphics[scale=0.45,angle=0]{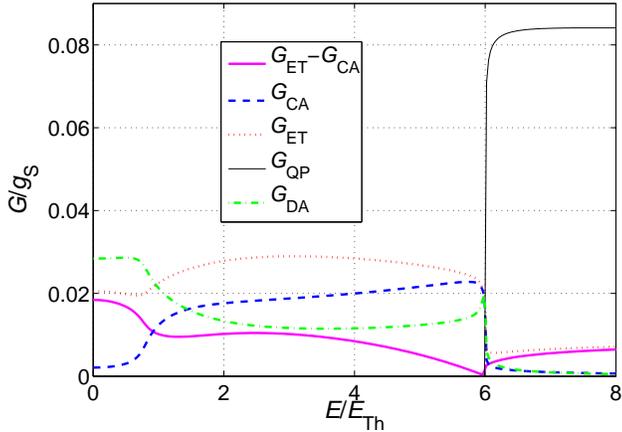}
  \caption{Conductances when N$_1$ and N$_2$ are connected by point
    contacts, and S by an interface of intermediate transparency
    $T_k^\text{(S)}=0.5$. The nonlocal conductance is largest for
    energy smaller than the Thouless energy, and has features at
    energy corresponding to the gap of the superconducting reservoir
    at $\Delta=6~E_\text{Th}$.}
  \label{fig:NpointSinter}
\end{figure}

The equations that determine $\theta$ must generally be solved
numerically. We have performed such calculations, and show in Fig.
\ref{fig:NpointSinter} the result for a system where connectors to
N$_1$ and N$_2$ are point contacts and the connector to S has
transparency $T^\text{(S)}_k=0.5$ for the conducting modes. Defining
$g_n=e^2\sum_kT^{(n)}_k/(\pi\hbar)$ we choose parameters
$g_j/g_\text{S}=0.1$ for $j=1,2$. There are now two energy scales in
the problem, $\Delta$ and $E_\text{Th}$, and we have chosen
$\Delta/E_\text{Th}=6$ in Fig. \ref{fig:NpointSinter}. The nonlocal
conductance is largest for energy smaller than the effective Thouless
energy, defined as $E_\text{Th}=\hbar
g_\text{S}/(2e^2\nu_0V_\text{c})$, and decreases in two steps at
$E_\text{Th}$ and $\Delta$ with increasing energy. The nonlocal
conductance above the gap corresponds to the normal state result
$\partial I_1/\partial V_2=g_1g_2/(g_1+g_2+g_\text{S})$ for
quasiparticle transport in a three terminal network. In this energy
range we also have a contribution from quasiparticle transfer into S.

\begin{figure}[h]
  \includegraphics[scale=0.45,angle=0]{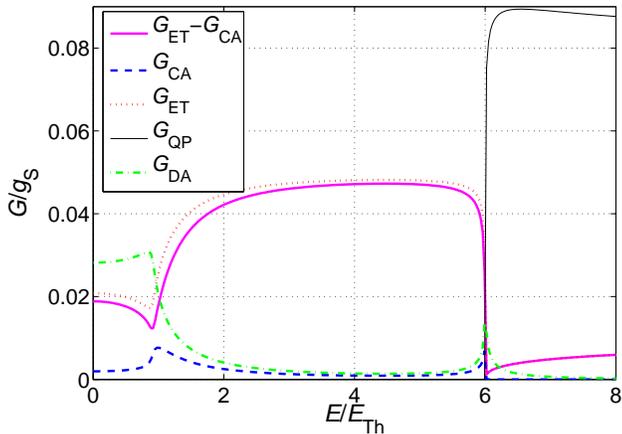}
  \caption{Conductances when N$_1$ and N$_2$ are connected by point
    contacts, and S by a tunnel barrier. In contrast to the the device
    where S was connected by an interface of intermediate transparency
    (Fig.  \ref{fig:NpointSinter}), the CA conductance below the gap
    is strongly suppressed in this case.}
  \label{fig:NpointStunnel}
\end{figure}

When the transmission probabilities of the interface to S are in the
tunneling limit, i.e., all $T_k^{(\text{S})}\ll 1$ the CA conductance
will be suppressed in comparison to the case shown in Fig.
\ref{fig:NpointSinter}. The ET conductance, on the other hand, is
enhanced by the reduced transmission of the contact to S. In Fig.
\ref{fig:NpointStunnel} we show the conductances for a device where
N$_1$ and N$_2$ are connected by point contacts, and S by a tunneling
barrier. The conductance of the tunneling barrier is the same as the
conductance of the contact to S in Fig. \ref{fig:NpointSinter}. For
energies between $E_\text{Th}$ and $\Delta$, we see that $G_\text{CA}$
is strongly suppressed. The total nonlocal conductance,
$G_\text{ET}-G_\text{CA}$, has a minimum in the subgap regime at
energy corresponding to $E_\text{Th}$, and is largest for an energy
close to $\Delta$. This is qualitatively different from the device
where S is connected by an interface of intermediate transparency
(Fig. \ref{fig:NpointSinter}), where the maximum nonlocal conductance
in the subgap regime is found at very small energy, and then decreases
with increasing energy due to increasing CA conductance.

\section{\label{sec:conc}Conclusion}
%*******************************************************************
In conclusion, we have studied nonlocal transport in a three-terminal
device with two normal-metal terminals and one superconducting
terminal. To this end we have applied the circuit theory of mesoscopic
transport. The connectors between the circuit elements are represented
by general expressions, relevant for a wide range of contacts.
Dephasing is taken into account, and gives rise to an effective
Thouless energy. We calculate the conductance for crossed Andreev
reflection, electron transfer between the normal-metal terminals, and
direct Andreev reflection and quasiparticle transport between one
normal-metal terminal and the superconducting terminal. The nonlocal
conductance is generally dominated by electron transfer in our model,
similar predictions were made in Refs.
\cite{melin:174509,morten:prb06}. We showed in Ref.
\cite{morten:prb06} that for this model, in the limit that there is no
resistance between the device and the superconducting terminal, our
results agree with Ref. \cite{Falci:epl01} and the total nonlocal
conductance vanishes. We numerically calculate the conductances for
experimentally relevant combinations of contacts to demonstrate the
appearance of two energy scales in the conductances: The effective
Thouless energy and the gap of the superconducting terminal. The
conductance for crossed Andreev reflection depends strongly on the
transparency of the interface to superconducting terminal as
demonstrated in our numerical calculations.

\section*{Acknowledgments}
This work was supported in part by The Research Council of Norway
through Grants No. 167498/V30, 162742/V00, 1534581/432, 1585181/143,
1585471/431, the DFG through SFB 513, the Landesstiftung
Baden-W\"{u}rttemberg, the National Science Foundation under Grant No.
PHY99-07949, and EU via project NMP2-CT-2003-505587 'SFINx'.

\bibliographystyle{aip}
\bibliography{/home/gudrun/janpette/artikkel/fs}

\end{document}